\documentclass[journal]{IEEEtran}
\ifCLASSINFOpdf
\else
\fi
\usepackage{listings}
\usepackage{xcolor}
\usepackage{rotating}
\usepackage{multicol}
\usepackage{multirow}
\usepackage{verbatim}

\definecolor{mGreen}{rgb}{0,0.6,0}
\definecolor{mGray}{rgb}{0.5,0.5,0.5}
\definecolor{mPurple}{rgb}{0.58,0,0.82}
\definecolor{backgroundColour}{rgb}{1,1,1}

\lstdefinestyle{CStyle}{
    backgroundcolor=\color{backgroundColour},   
    commentstyle=\color{mGreen},
    keywordstyle=\color{magenta},
    numberstyle=\tiny\color{mGray},
    stringstyle=\color{mPurple},
    basicstyle=\footnotesize,
    breakatwhitespace=false,         
    breaklines=true,                 
    captionpos=b,                    
    keepspaces=true,                 
    numbers=left,                    
    numbersep=5pt,                  
    showspaces=false,                
    showstringspaces=false,
    showtabs=false,                  
    tabsize=2,
    language=C
}

\begin{document}
%
\title{An Embedded RISC-V Core with Fast Modular Multiplication}
%
%
%

\author{\"Omer Faruk Irmak, Arda Yurdakul
\thanks{ \"Omer Faruk IRMAK was with the Department
of Computer Engineering, Bo\u{g}azi\c{c}i University, Bebek 34342, Istanbul, Turkey\protect\\
E-mail: omer.irmak@boun.edu.tr}
\thanks{Arda Yurdakul was with the Department
of Computer Engineering, Bo\u{g}azi\c{c}i University, Bebek 34342, Istanbul, Turkey\protect\\
E-mail: yurdakul@boun.edu.tr}
\thanks{Manuscript received September 30, 2020.}}

\maketitle

\begin{abstract}
 One of the biggest concerns in IoT is privacy and security. Encryption and authentication need big power budgets, which battery-operated IoT end-nodes do not have. Hardware accelerators designed for specific cryptographic operations provide little to no flexibility for future updates. Custom instruction solutions are smaller in area and provide more flexibility for new methods to be implemented. One drawback of custom instructions is that the processor has to wait for the operation to finish. Eventually, the response time of the device to real-time events gets longer. In this work, we propose a processor with an extended custom instruction for modular multiplication, which blocks the processor, typically, two cycles for any size of modular multiplication when used in Partial Execution mode. We adopted embedded and compressed extensions of RISC-V for our proof-of-concept CPU. Our design is benchmarked on recent cryptographic algorithms in the field of elliptic-curve cryptography. Our CPU with 128-bit modular multiplication operates at 136MHz on ASIC and 81MHz on FPGA. It achieves up to 13x speed up on software implementations while reducing overall power consumption by up to 95\% with 41\% average area overhead over our base architecture.
\end{abstract}


%
\IEEEpeerreviewmaketitle

\section{Introduction}

\IEEEPARstart{I}{oT} market has been one of the driving forces of embedded hardware. Key enabler of IoT is cheap and capable hardware. There are multiple efforts, both in academia and industry, that are aiming to bring costs lower while making hardware more efficient in the tasks it is designed to perform. IoT end-node hardware should be secure and designed with power consumption in mind. Therefore, there are efforts on both designing new lightweight algorithms \cite{nist_lw_crypto} that suit better to less powerful processors and designing specialized hardware that tackles the heavy operations more efficiently \cite{6665020}.

Custom instructions can be utilized for accelerating cryptographic operations. Fundamental and complex operations in cryptography can be mapped to custom instructions and implemented in hardware with fewer resources compared to full custom accelerators. This makes using the same hardware for different algorithms possible as custom instructions can be utilized in the realization of any algorithm. If the current algorithm turns out to be vulnerable, different solutions can be implemented via a software update without a significant performance penalty.

In this work we have designed a microprocessor core with its ISA extended with a custom instruction for Montgomery multiplication. Modular multiplication is highly utilized in public key cryptography. Our proposed custom instruction implementation can be executed both atomically and partially in short iterations, therefore does not degrade system response time. We implemented Embedded and Compressed extensions of RISC-V (RV32EC) \cite{riscv_2019} as the base ISA of our proof-of-concept CPU. Design is benchmarked with operations on various cryptographic elliptic curves. Synthesis is done for both FPGA and ASIC targets to collect area and power consumption metric. Our contributions can be summarized as follows;

\begin{itemize}
    \item Propose a multiprecision MMUL custom instruction
    \item Propose a method to partition runtime of a long-latency custom instruction to increase responsiveness of the CPU to external events
    \item Analyze different RISC-V instruction encodings available to be used for custom instructions
\end{itemize}

\noindent In the literature, there are plenty of studies for adding custom instructions to RISC-V. Yet, none of them studies the effects of blocking the processor with a custom instruction or effects of the encoding within our knowledge.

\section{Montgomery Multiplication Instruction for RISC-V ISA}

Modular multiplication is the operation of \begin{math} P = (A * B)\ mod\ N \end{math} . One of the key efficient algorithms in this area is Montgomery Multiplication \cite{Montgomery85}. For operands with length of $n$ in bits, Montgomery multiplication calculates \begin{math} MMUL(A, B, N) = (A * B * R\textsuperscript{-1})\ mod\ N \end{math} where \begin{math} R = 2\textsuperscript{(2$n$)}\ mod\ N \end{math}, \begin{math} 2\textsuperscript{$n$-1} < N < 2\textsuperscript{$n$} \end{math} and \begin{math} gcd(R, N) = 1 \end{math}. We chose the Radix-2 Montgomery Multiplication (R2MM) algorithm \cite{1228516} for the implementation. R2MM is suitable for a simple hardware implementation as it is composed of additions and shifts.

\begin{figure}[h]
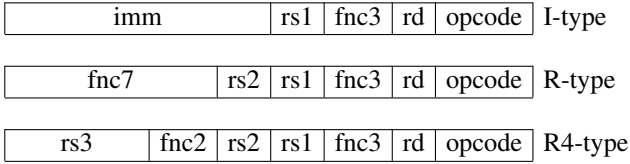

\begin{center}
\setlength{\tabcolsep}{4pt}
\begin{tabular}{p{0.7
in}@{}p{0.5in}@{}p{0.7in}@{}p{0.7in}@{}p{0.6in}@{}p{0.7in}@{}p{0.8in}l}
\cline{1-7}
\multicolumn{3}{|c|}{imm} &
\multicolumn{1}{c|}{rs1} &
\multicolumn{1}{c|}{fnc3} &
\multicolumn{1}{c|}{rd} &
\multicolumn{1}{c|}{opcode} &
I-type \\
\cline{1-7}
\\
\cline{1-7}
\multicolumn{2}{|c|}{fnc7} &
\multicolumn{1}{c|}{rs2} &
\multicolumn{1}{c|}{rs1} &
\multicolumn{1}{c|}{fnc3} &
\multicolumn{1}{c|}{rd} &
\multicolumn{1}{c|}{opcode} &
R-type \\
\cline{1-7}
\\
\cline{1-7}
\multicolumn{1}{|c|}{rs3} &
\multicolumn{1}{c|}{fnc2} &
\multicolumn{1}{c|}{rs2} &
\multicolumn{1}{c|}{rs1} &
\multicolumn{1}{c|}{fnc3} &
\multicolumn{1}{c|}{rd} &
\multicolumn{1}{c|}{opcode} &
R4-type \\
\cline{1-7}
\end{tabular}
\end{center}
\caption{Candidate RISC-V instruction formats}
\label{fig:baseinstformats}
\end{figure}

In RISC-V, different instruction formats have already been defined. Some of them can be seen in Figure \ref{fig:baseinstformats}. Regardless of the instruction encoding, we decided MMUL instruction to work on memory addresses unlike any instruction in RISC-V specification, which strictly works on register values. When it comes to multiprecision operations, defining a unified interface on memory addresses is more performant. The key point that has to be made clear is the layout of in memory. Constraining how operands should be arranged may result in lower performance as it may require application code to rearrange operands in memory. MMUL requires 3 memory addresses for the inputs and a single memory address for the output. Length of the operands must be encoded in the instruction for flexibility. Operand length may be limited by the hardware implementation of the MMUL instruction. In our reference design, maximum operand length is a hardware constraint that is defined at the synthesis phase. 

\begin{figure}[h!]
\centering
\includegraphics[scale=0.5]{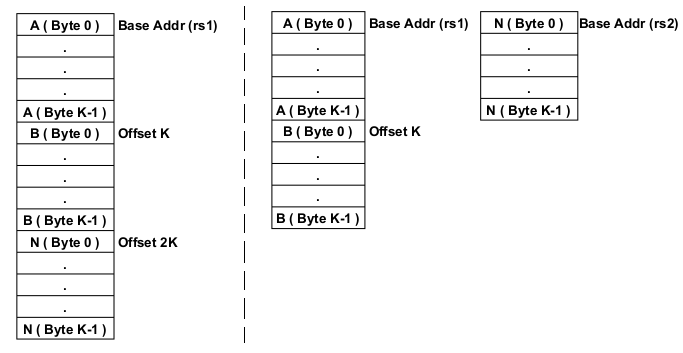}
\caption{Memory layout for I-type (left) and R-type (right) MMUL}
\label{fig:irtype_mem}
\end{figure}

If application can guarantee that all operands will be in a certain offset from a base address in memory as shown in Figure \ref{fig:irtype_mem}, a single memory address stored in rs1 is enough for the input operands. Thus I-type instruction format can be used. Fields fnc3 and imm provide 15 bits in the instruction to be used for encoding length, which can be encoded in bits to give a maximum of 32768 bit operands.

Likewise, if multiplicand and multiplier are guaranteed to be always in fixed positions relative to each other but modulus may be in random addresses as shown in Figure \ref{fig:irtype_mem}; R-type format can be used. Two source registers, rs1 and rs2, would be used as base addresses. The fnc3 and fnc7 fields give 10 bits of space which enables, if length is encoded in bits, a maximum of 1024 bit operands. 

\begin{figure}[h!]
\centering
\includegraphics[scale=0.35]{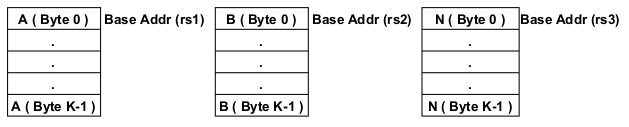}
\caption{Memory layout for R4-type MMUL}
\label{fig:r4type_mem}
\end{figure}

Lastly, for the best performance in all cases, if R4-type format is used, all operands can be in their independent addresses stored in rs1, rs2 and rs3 as shown in Figure \ref{fig:r4type_mem}. This format leaves only 5 bits (fnc3 and fnc2) which is not enough for length to be encoded in bits. Encoding operand length in words is another option which makes 1024 bit (2\textsuperscript{5} * 32) length operands for RV32 and 2048 bit (2\textsuperscript{5} * 64) operands for RV64 possible. 

In this work, we decided to use R4-type instruction format because it imposes no memory layout restrictions. Using GCC directive $.insn$ \cite{insn_directive} in this decision process sped up the development.

\begin{figure}[h]
\centering
\includegraphics[scale=0.35]{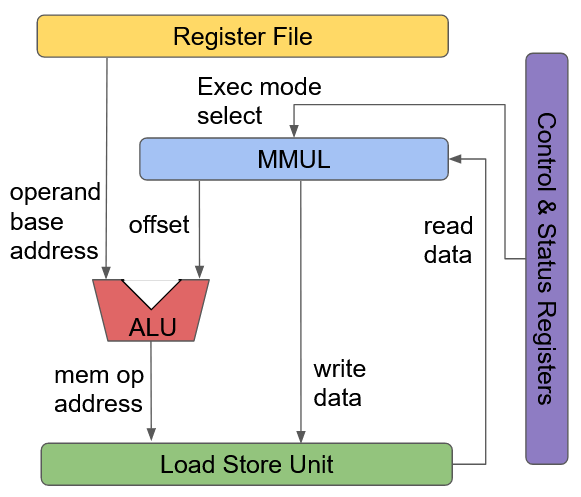}
\caption{Integration of MMUL in datapath}
\label{fig:mm_integ}
\end{figure}

As it can be seen in Figure \ref{fig:mm_integ}, MMUL is coupled with the datapath of the processor. Addresses of the operands are read directly from their respective registers of the Register File and fed to the ALU in the datapath. Memory address to be worked on is calculated in the ALU by adding the offset value supplied by the MMUL to the base address read from the Register File. LSU is triggered by MMUL module to load or store from the calculated address. Operands are loaded at the start of the execution and stored in MMUL module during the entire operation. All execution is controlled by MMUL itself.

\section{Partial Execution Mode}

R2MM \cite{1228516} has a loop with $n$ iterations and one final subtraction after exiting the loop. Our implementation takes 2 clock cycles for each loop iteration and one last cycle for the subtraction. So in total, one MMUL operation takes $2n + 1$ clock cycles for calculations, $3*WORDS$ memory load operations for fetching the three operands and $WORDS$ memory write operations for writing back the result where $WORDS = ceil(n / WORD\_SIZE)$. 

As the instructions are atomic, during MMUL operation, processor will be unresponsive to any event that may happen. For some applications this may be problematic because of the real time constraints they have. To remedy this we can move the loop in our algorithm from hardware to software, allowing our processor to service interrupts in between loop iterations.

\begin{figure}[!h]
\centering
\includegraphics[scale=0.25]{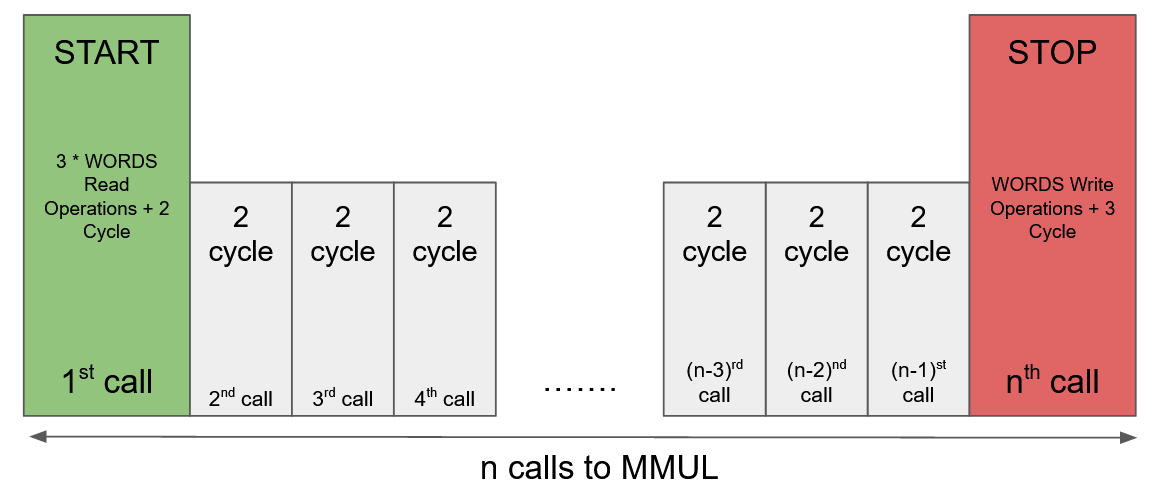}
\caption{MMUL partial execution time diagram}
\label{fig:mm_time_diag}
\end{figure}

\begin{figure*}[htb]
\centering
\includegraphics[scale=0.55]{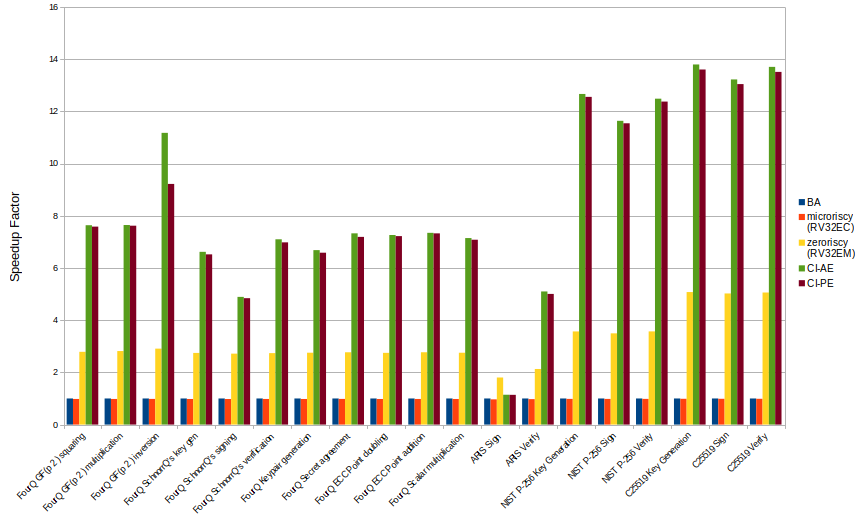}
\caption{Benchmark Speedups, Base RV32EC Architecture (BA), Custom Instruction with Atomic Execution (CI-AE), Custom Instruction with Partial Execution (CI-PE)}
\label{fig:perf_results}
\end{figure*}

To achieve this behaviour, which we call \textit{partial execution}, our implementation has a special Control and Status Register (CSR). If partial execution is enabled by a write with $csrwr$ instruction to this register, which is directly connected to "Execution Mode Select" signal in Figure \ref{fig:mm_integ}, current MMUL instruction is retired after each bit is processed. Application code has to execute another MMUL instruction for each bit of operands, ie. n calls to the MMUL for an n-bit Montgomery multiplication operation, as shown in Figure \ref{fig:mm_time_diag}. First call to the MMUL does the memory load operations, while last call writes back the result. In this case, maximum latency of MMUL instruction drops to either $3*WORDS$ memory load operations + 2 cycles or $WORDS$ memory write operations + 3 cycles depending on the memory operation latencies. Performance penalty of this, which will be later presented, is minimal when used with loop unrolling.

\section{Analysis}

\subsection{Base Architecture}

To set a baseline for our work, we designed an in-order 2-stage RV32EC core with minimal area while maintaining comparable level of performance. Coremark and Dhrystone are run both on microriscy \cite{8106976} and our core using the same memory modules. While our core scored 0.905 in Coremark and 1805 in Dhrystone, microriscy \cite{8106976} scored 0.878 and 1644, respectively. Even though our core is slightly faster than microriscy, they can be considered equal in terms of performance.

\subsection{Benchmarks}

Our design is benchmarked with multiple ECC curves. Software implementations of FourQ (128-bit)\cite{fourq_paper}, NIST P-256 (256-bit)\cite{Hess2000SEC2R}, Curve25519 (256-bit)\cite{10.1007/11745853_14} and ARIS (an authentication scheme based on FourQ) \cite{aris} are run on our processor and microriscy/zeroriscy \cite{8106976} cores from PULP as the reference designs. Later, modular multiplication and squaring implementations are replaced with a sequence of MMUL instructions and run on our modified core. No modifications are made to any other part of the code.

In Figure \ref{fig:perf_results}, speedup of ECC benchmarks can be seen. Full software benchmarks are run on microriscy/zeroriscy \cite{8106976} and our base architecture (BA) as the control group. Tests where modular multiplication operation is implemented with our custom instruction are labeled Custom Instruction with Atomic Execution (CI-AE) and Custom Instruction with Partial Execution (CI-PE). There is a significant speed up in all curve operations. This speed up contributes to lowering total energy consumption. Our experimental setup uses a memory unit with single cycle read latency. Longer latency memories like EEPROM, FLASH are widely used as instruction memories. If a longer latency instruction memory was used in benchmarks, results would be even more in favour of our implementation. MMUL instruction takes multiple cycles thus allowing a new instruction to be fetched before it finishes, therefore CPU is less likely to be stalled waiting for new instructions.

Performance penalty of partial execution is negligible when paired with loop unrolling. Depending on the compiler output, if not interrupted, a Montgomery multiplication can be executed in the same amount of time as atomic execution.

For power consumption analysis, FPGA tools are used. Design is synthesized on a Xilinx XC7Z020-1 FPGA and activity data is gather with Xilinx's development environment, Vivado. Activity data is then used to increase dynamic power estimation accuracy.

\begin{table}[h!]
\centering
\caption{Average Power Consumption (W) During A Modular Multiplication}
\label{tab:powers}
\begin{tabular}{l|l|l|l}
\centering
 & \textbf{Static} & \textbf{Dynamic} & \textbf{Total} \\
\hline
BA                & 0.107     & 0.154     & 0.261   \\ 
CI-AE     & 0.105     & 0.064     & 0.170   \\ 
CI-PE       & 0.106     & 0.120     & 0.226   \\
\end{tabular}
\end{table}

Power consumption of different configurations can be seen in Table \ref{tab:powers}. While static power consumption shows only small changes, dynamic power goes down significantly. This can be explained with the power consumption per design block during fully software and with custom instruction runs of the benchmark (Table \ref{tab:per_module_powers}). When executing solely standard instructions, as in BA column of Table \ref{tab:per_module_powers}, every module of the CPU works synchronously. When MMUL instruction is in progress, rest of the CPU is idle. Biggest gain comes from the fetch stage because only four instruction fetches are needed per modular multiplication with our custom instruction and it is the biggest module in the design.

\begin{table}[h]
\centering
\caption{Average Dynamic Power Consumption Per Module (W)}
\label{tab:per_module_powers}
\begin{tabular}{l|l|l|l}
\centering
 & \textbf{BA} &  \textbf{CI-AE} &   \textbf{CI-PE}   \\
\hline
Fetch Stage     & 0.058 &  0.002  &  0.026 \\
Decoder         & 0.014 &  0.001  &  0.006 \\
ALU             & 0.031 &  0.001  &  0.008 \\
Register File   & 0.012 &  0.002  &  0.003 \\
MMUL            & 0     &  0.054  &  0.053  \\

\end{tabular}
\end{table}

Both average power consumption and execution time go down in our implementation. Naturally, product of these two metrics follow this trend as well. Normalized energy consumption values can be seen in Table \ref{tab:normal_powers}. For FourQ, a 128-bit curve, roughly 90\% of energy is saved while P-256/C25519, 256-bit curves, savings go up to 95\%. As the prime that curves use gets bigger, performance increases and this results in higher energy savings. R2MM scales better for larger operands with an O(n) time complexity \cite{1228516} \cite{Karatsuba1963MultiplicationOM}.

\begin{table}[h]
\centering
\caption{Normalized Energy Consumption (Power x Clock Cycles)}
\label{tab:normal_powers}
\begin{tabular}{l|l|l|l|l}
\centering
 & & \textbf{BA} & \textbf{CI-AE} & \textbf{CI-PE} \\
\multirow{3}{*}{\begin{turn}{90} FourQ \end{turn}} & 
KeyGen & 1  & 0.10 & 0.13  \\ 
& Sign & 1  & 0.13  & 0.18  \\ 
& Verify & 1  & 0.09  & 0.12  \\
\hline
\multirow{3}{*}{\begin{turn}{90} P256\end{turn}} & 
KeyGen & 1  & 0.05  & 0.07  \\ 
& Sign & 1  & 0.06  & 0.08  \\ 
& Verify & 1  & 0.05  & 0.07  \\ 
\hline
\multirow{3}{*}{\begin{turn}{90} C25519\end{turn}} & 
KeyGen & 1  & 0.05   & 0.07  \\ 
& Sign & 1  & 0.05   & 0.07  \\ 
& Verify & 1  & 0.05  & 0.06  \\ 
\end{tabular}
\end{table}

Although MMUL itself is fairly small, it adds 33\% area overhead (from 487 to 649 slices) to our base architecture and operating frequency goes down by 9\% (from 89MHz to 81Mhz) in FPGA synthesis. Using the TSMC OSU 0.18um technology, ASIC synthesis shows 49\% area overhead (from 872 FF and 8106 gates to 1305 FF and 12105 gates) and 8\% decrease  (from 148MHz to 136Mhz) in operating frequency. Depending on the requirements of the application, a different implementation of Montgomery multiplication may be used for the required balance between performance gain and area overhead. It is debated \cite{iotecc} that ECC is too complex to be used on IoT devices, yet even new lightweight algorithms introduce similar overheads when accelerated in hardware. To give a comparison, Tehrani et. al. \cite{tehrani} accelerate Lightweight Block Ciphers on a RV32I platform. On average, their work introduces 58\% area overhead.

\section{Conclusion}

In this paper, we proposed a microprocessor core with a custom instruction for Montgomery multiplication.  Radix-2 Montgomery multiplication is implemented as an instruction. For better system response times, a partial execution scheme is proposed, enabling the instruction to be completed in multiple short-latency iterations. The resulting hardware is realized on FPGA and as an ASIC. 136 MHz clock speed on ASIC and 81 MHz clock speed on FPGA with 128-bit MMUL module is achieved. It achieves up to 13x speed up on various cryptographic curves compared to software implementations while reducing overall power consumption by up to 95\%.

\ifCLASSOPTIONcaptionsoff
  \newpage
\fi



%
\bibliographystyle{IEEEtran}
\bibliography{references}

\begin{thebibliography}{10}
\providecommand{\url}[1]{#1}
\csname url@samestyle\endcsname
\providecommand{\newblock}{\relax}
\providecommand{\bibinfo}[2]{#2}
\providecommand{\BIBentrySTDinterwordspacing}{\spaceskip=0pt\relax}
\providecommand{\BIBentryALTinterwordstretchfactor}{4}
\providecommand{\BIBentryALTinterwordspacing}{\spaceskip=\fontdimen2\font plus
\BIBentryALTinterwordstretchfactor\fontdimen3\font minus
  \fontdimen4\font\relax}
\providecommand{\BIBforeignlanguage}[2]{{%
\expandafter\ifx\csname l@#1\endcsname\relax
\typeout{** WARNING: IEEEtran.bst: No hyphenation pattern has been}%
\typeout{** loaded for the language `#1'. Using the pattern for}%
\typeout{** the default language instead.}%
\else
\language=\csname l@#1\endcsname
\fi
#2}}
\providecommand{\BIBdecl}{\relax}
\BIBdecl

\bibitem{nist_lw_crypto}
\BIBentryALTinterwordspacing
C.~S. Division, I.~T. Laboratory, N.~I. of~Standards, Technology, and
  D.~of~Commerce, ``Lightweight cryptography.'' [Online]. Available:
  \url{https://csrc.nist.gov/projects/lightweight-cryptography}
\BIBentrySTDinterwordspacing

\bibitem{6665020}
B.~{Blaner}, B.~{Abali}, B.~M. {Bass}, S.~{Chari}, R.~{Kalla}, S.~{Kunkel},
  K.~{Lauricella}, R.~{Leavens}, J.~J. {Reilly}, and P.~A. {Sandon}, ``Ibm
  power7+ processor on-chip accelerators for cryptography and active memory
  expansion,'' \emph{IBM Journal of Research and Development}, vol.~57, no.~6,
  pp. 3:1--3:16, Nov 2013.

\bibitem{riscv_2019}
\BIBentryALTinterwordspacing
``Offical riscv foundation website,'' Oct 2019. [Online]. Available:
  \url{https://riscv.org/}
\BIBentrySTDinterwordspacing

\bibitem{Montgomery85}
P.~L. Montgomery, ``Modular multiplication without trial division,''
  \emph{Mathematics of Computation}, vol.~44, no. 170, pp. 519--521, 1985.

\bibitem{1228516}
A.~F. {Tenca} and C.~K. {Koc}, ``A scalable architecture for modular
  multiplication based on montgomery's algorithm,'' \emph{IEEE Transactions on
  Computers}, vol.~52, no.~9, pp. 1215--1221, Sep. 2003.

\bibitem{insn_directive}
\BIBentryALTinterwordspacing
 [Online]. Available:
  \url{https://embarc.org/man-pages/as/RISC\_002dV\_002dFormats.html}
\BIBentrySTDinterwordspacing

\bibitem{8106976}
P.~{Davide Schiavone}, F.~{Conti}, D.~{Rossi}, M.~{Gautschi}, A.~{Pullini},
  E.~{Flamand}, and L.~{Benini}, ``Slow and steady wins the race? a comparison
  of ultra-low-power risc-v cores for internet-of-things applications,'' in
  \emph{2017 27th International Symposium on Power and Timing Modeling,
  Optimization and Simulation (PATMOS)}, Sep. 2017, pp. 1--8.

\bibitem{fourq_paper}
C.~Costello and P.~Longa, ``Fourq: Four-dimensional decompositions on a q-curve
  over the mersenne prime,'' in \emph{ASIACRYPT}, 2015.

\bibitem{Hess2000SEC2R}
P.~Hess, ``Sec 2: Recommended elliptic curve domain parameters,'' 2000.

\bibitem{10.1007/11745853_14}
D.~J. Bernstein, ``Curve25519: New diffie-hellman speed records,'' in
  \emph{Public Key Cryptography - PKC 2006}, M.~Yung, Y.~Dodis, A.~Kiayias, and
  T.~Malkin, Eds.\hskip 1em plus 0.5em minus 0.4em\relax Berlin, Heidelberg:
  Springer Berlin Heidelberg, 2006, pp. 207--228.

\bibitem{aris}
R.~{Behnia}, M.~O. {Ozmen}, and A.~A. {Yavuz}, ``Aris: Authentication for
  real-time iot systems,'' in \emph{ICC 2019 - 2019 IEEE International
  Conference on Communications (ICC)}, 2019, pp. 1--6.

\bibitem{Karatsuba1963MultiplicationOM}
A.~Karatsuba and Y.~P. Ofman, ``Multiplication of many-digital numbers by
  automatic computers,'' in \emph{Doklady Akademii Nauk SS}, vol.~14, no. 145,
  1963, pp. 293--294.

\bibitem{iotecc}
M.~Samaila, J.~Sequeiros, T.~Simões, M.~Freire, and P.~Inácio,
  ``Iot-harpseca: A framework and roadmap for secure design and development of
  devices and applications in the iot space,'' \emph{IEEE Access}, vol.~PP, pp.
  1--1, 01 2020.

\bibitem{tehrani}
E.~{Tehrani}, T.~{Graba}, A.~S. {Merabet}, S.~{Guilley}, and J.~{Danger},
  ``Classification of lightweight block ciphers for specific processor
  accelerated implementations,'' in \emph{2019 26th IEEE International
  Conference on Electronics, Circuits and Systems (ICECS)}, 2019, pp. 747--750.

\end{thebibliography}

%




\end{document}